\documentclass[aps,twocolumn,showpacs,floats,superscriptaddress]{revtex4}
\usepackage{dcolumn}
\usepackage{amsmath, amssymb, bm, graphicx}
\def\beq{\begin{equation}}
\def\eeq{\end{equation}}

\begin{document}

\author{A.B. Kuklov}
\affiliation{Department of Engineering Science and Physics, CUNY,
Staten Island, NY 10314}
\author{M. Matsumoto}
\affiliation{Theoretische Physik, ETH Zurich, 8093 Zurich, Switzerland}
\affiliation{Department of Physics, University of California,
Davis, CA 95616}
\author{N.V. Prokof'ev}
\affiliation{Theoretische Physik, ETH Zurich, 8093 Zurich, Switzerland}
\affiliation{Department of Physics, University of
Massachusetts, Amherst, MA 01003, USA}
 \affiliation{Russian Research Center
``Kurchatov Institute'', 123182 Moscow, Russia}
\author{B.V. Svistunov}
\affiliation{Department of Physics, University of Massachusetts,
Amherst, MA 01003, USA} \affiliation{Russian Research Center
``Kurchatov Institute'', 123182 Moscow, Russia}
\author{M. Troyer}
\affiliation{Theoretische Physik, ETH Zurich, 8093 Zurich, Switzerland}

\title{Search for Deconfined Criticality: SU(2) D\'{e}j\`{a} Vu}

\date{\today}
\begin{abstract}
Monte Carlo simulations of the SU(2)-symmetric deconfined critical point action
reveal strong violations of scale invariance for the deconfinement transition. We find
compelling evidence that the generic runaway renormalization flow of the
gauge coupling is to a  weak first order transition, similar to the case of U(1)$\times$U(1) symmetry. Our
results imply that recent numeric studies of the N\`{e}el antiferromagnet
to valence bond solid quantum phase transition in SU(2)-symmetric
models were not accurate enough in determining the nature of the transition.
\end{abstract}

\pacs{05.30.-d, 75.10.-b, 05.50.+q}

\maketitle

Within the standard Ginzburg-Landau-Wilson description of critical
phenomena a direct transition between states which break different
symmetries is expected to be of first-order.
The existence of a generic line
of deconfined critical points (DCP) proposed in
Refs.~\cite{Motrunich,dcp1,dcp2} --- an exotic second-order phase transition
between two competing orders --- remains one of the most intriguing
and controversial topics in the modern theory of phase transitions.
In particular, the DCP theory makes the prediction that certain types of superfluid
to solid and the N\`{e}el antiferromagnet to valence bond solid (VBS)
quantum phase transitions in 2D lattice systems can be  continuous. Remarkably, the new criticality is in the
same universality class as a 3D system of $N=2$ identical complex-valued
classical fields coupled to a gauge vector field (referred to as the DCP action below). This makes the DCP theory
relevant also for the superfluid to normal liquid
transition in symmetric two-component superconductors \cite{Babaev}.

An intrinsic difficulty in understanding properties of the $N$-component
DCP action is its runaway renormalization flow to strong coupling at large scales
and the absence of perturbative fixed points for realistic $N$ \cite{Halperin,Sachdev}.
One may only speculate that the value of $N$ might be of little importance since
the possibility of the continuous transition for $N=1$ is guaranteed by the exact duality
mapping between the inverted-XY and XY-universality classes \cite{duality} and for
$N\to \infty$  it follows from the large-$N$ expansion for $N$ of the order of a hundred.
However, there are no exact analytic results either showing that in a
two-component system there exists a generic line of second-order phase transitions,
or proving that the second-order phase transition is fundamentally impossible.
The problem of deconfined criticaly for the most interesting
case of $N=2$ thus has to be resolved by numerical simulations.

The initial effort was focused on models of the superfluid to solid quantum phase transitions
and U(1)$\times$U(1)-symmetric DCP actions \cite{Sandvik2002,Motrunich}.
First  claims of deconfined criticality were confronted with
the observation of weak first-order transitions in other models \cite{weak_first}.
While presenting a particular model featuring a first order phase transition does not prove
the impossibility of a continuous DCP yet, it does raise a warning flag.  One needs to pay special attention to any signatures of violation of the scale invariance which  may be indicative of a runaway flow to a first-order transition
even when all other quantities appear to change continuously
due to limited system sizes available in simulations \cite{prog_theor}.
The flowgram method \cite{flowgram} was developed as a generic tool for monitoring such
runaways flow to strong
coupling and was
used to prove the generic first-order nature of the deconfinement transition in the
U(1)$\times$U(1)-symmetric DCP action. A subsequent refined analysis resulted in the 
reconsideration of the original claims in favor of a discontinuous transition for all known models
\cite{Sandvik2006,Sudbo2006}.

Recently the SU(2)-symmetric case has been studied in a series of papers
\cite{Sandvik2007,Melko2008,MV} and an exciting observation of a
continuous DCP point was reported. However, the story seems to
repeat itself since renormalization flows for the $J$-$Q$ model studied in
Refs.~\cite{Sandvik2007,Melko2008} were shown to be in violation of
scale invariance and, possibly,  indicative of the first-order transition \cite{Wiese}. 
In this Letter we show that a runaway flow
to strong coupling and a first order transition is a generic feature
of all SU(2)-symmetric DCP models 
analogous to the U(1)$\times$U(1) case \cite{su2}.

For our simulations we consider the lattice version of the SU(2)-symmetric NCCP$^1$ model
\cite{dcp1,dcp2} and map it onto the two-component $J$-current model.
The DCP action for two spinon fields $z_a, \, a=1,2$ on a three-dimensional simple cubic
lattice is defined as
\begin{eqnarray}
S&=&  - \sum_{<ij>, a} t(z^*_{ai}z^{\:}_{aj}e^{iA_{<ij>}}+c.c)
\nonumber \\
& +& \frac{1}{8g} \sum_{\Box} (\nabla \times A)^2\; ; \quad
\sum_{a}|z_{ai}|^2 =1 \;, \label{dcp_2} \label{dcp}
\end{eqnarray}
where $\langle ij\rangle$ runs over nearest neighbor pair of sites $i,j$,
the gauge field $A_{<ij>}$ is defined on the bonds,
and $\nabla \times A$ is a short-hand notation for the lattice curl-operator.
The mapping to the $J$-current model starts from the partition function $Z=\int Dz Dz^*
DA\exp(-S)$ and a Taylor expansion of the exponentials $\exp \{ t
z^*_{ai}z^{\:}_{aj}e^{iA_{<ij>}} \}$ and $\exp \{ t
z^*_{aj}z^{\:}_{ai}e^{-iA_{<ij>}} \}$ on all bonds. One can then perform an
explicit Gaussian integration over $A_{<ij>},\, z_{ai}$ and arrive at a
formulation in terms of integer non-negative bond currents $J^{(a)}_{i,\mu }$.
We use $\mu=\pm 1,\pm 2,\pm 3$ to label the directions of bonds
going out of a given site the corresponding unit vectors are denoted by $\hat{\mu}$.
These $J$-currents obey the conservation laws:
\begin{equation}
\sum_{\mu } I^{(a)}_{i,\mu
}=0,\,\mbox{ with  }\,I^{(a)}_{i,\mu }\equiv J^{(a)}_{i,\mu }-J^{(a)}_{i+\hat{\mu},
-\mu }.
\label{Kir}
\end{equation}
The final expression for the partition function reads
\begin{equation}
Z=\sum_{\{
J \} } {\cal Q}_{\rm site}\: {\cal Q}_{\rm bond} \: \exp(-H_J),
\end{equation}
 where
\begin{eqnarray}
H_J&=&\frac{g}{2} \sum_{ i,j;\, a,b;\, \mu=1,2,3}
I^{(a)}_{i,\mu }\, V_{ij} \, I^{(b)}_{j,\mu }
\label{H_J}\\
{\cal Q}_{\rm site}&=&\prod_i  \frac{{\cal N}^{(1)}_i! \, {\cal
N}^{(2)}_i!}{(1+{\cal N}^{(1)}_i+ {\cal N}^{(2)}_i)!},\quad
 {\cal N}^{(a)}_i=\frac{1}{2} \sum_{\mu} J^{(a)}_{i,\mu}
\nonumber \\
{\cal Q}_{\rm bond}&=&\prod_{i,a, \mu } \frac{t^{J^{(a)}_{i,\mu}}}{J^{(a)}_{i,\mu }!}
  \;, \nonumber
\end{eqnarray}
The long-range interaction $V_{ij}$ depends on the
distance $r_{ij}$ between the sites $i$ and $j$. Its Fourier transform
is given by $V_{\bf q} = 1/\sum_{\mu=1,2,3 } \sin^2 (q_{\mu}/2)$
and implies an asymptotic behavior $V \sim 1/r_{ij}$ at large
distances.

This formulation allows efficient Monte Carlo
simulations using a worm algorithm for the two-component system \cite{flowgram}. For the flowgram analysis we measure the mean square fluctuations of the winding
numbers $\langle W^2_{a,\mu}\rangle\equiv \langle W^2_{a,-\mu}\rangle$
of the conserved currents $ I^{(a)}_{i,\mu }$, or, equivalently, $\rho_\pm=\sum_\mu \langle (W_{1,\mu}\pm W_{2,\mu})^2\rangle/L \equiv \langle (W_{\pm}^2\rangle/L$.
In particular, we focused on the gauge invariant superfluid stiffness,
$\rho_-$ measuring the  response to a twist of the phase of the product $z^*_1z_2$.

\begin{figure}
\centerline{\includegraphics[width=\columnwidth, angle=0]{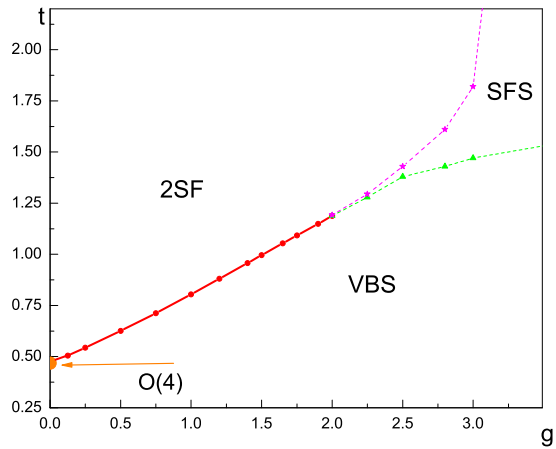}}
\caption{(Color online) Phase diagram of the SU(2)-symmetric DCP action~(\ref{dcp}).
I order transitions VBS-2SF are shown as solid red line up to the bicritical point $g_{bc}\approx 2.0$.}
\label{fig:pd}
\end{figure}

Similar to the U(1)$\times$U(1) case \cite{flowgram}, the NCCP$^{1}$ model features
three phases, Fig. \ref{fig:pd},
characterized by the following order parameters:
\begin{enumerate}
\item[VBS:]
an insulator with $\langle z_{ai}\rangle=0$
and, accordingly, $\langle \rho_+\rangle=\langle \rho_-\rangle=0$.
\item[2SF:]
two-component superfluid (2SF) with $\langle z_{ai}\rangle\neq 0$,
$\langle\rho+\rangle \neq 0$ and $\langle\rho_-\rangle \neq 0$.
\item[SFS:] supesolid (a paired phase \cite{note})  
 with $\langle z_{ai}\rangle=0,\, \langle
z^*_{1i}z_{2j}\rangle \neq 0$,  $\rho_+=0$ and $\rho_-\neq 0$.
\end{enumerate}
The point $g=0$ and $t\approx 0.468$ features a continuous transition
in the O(4) universality class. The relevant part of the phase diagram
is the region of small $g$ close to this O(4) point, far away from the bicritical point
$g_{bc}\approx 2.0$ where SFS phase intervenes between the
VBS and 2SF phases. The corresponding direct VBS-2SF
transition has been proposed to be a deconfined critical line (DCP line) \cite{dcp1,dcp2}.

The key idea of the flowgram method \cite{flowgram} is to demonstrate that the universal large-scale
behavior at $g\to 0$ is identical to that at some finite coupling $g=g_{\rm coll}$
where the nature of the transition can be easily revealed. The procedure is as follows:
\begin{enumerate}
\item[(i)] Introduce a definition of the critical point for a finite-size system
of linear size $L$ consistent with the thermodynamic limit and insensitive
to the order of the transition. In our model we used the same definition as in Ref.~\cite{flowgram}.
Specifically, for any given $g$ and $L$ we adjusted $t$ so that the ratio of statistical weights of configurations with and without windings 
was equal to $7.5$.
\
\item[(ii)] At the transition point, calculate a quantity $R(L,g)$ that is
supposed to be scale-invariant for a continuous phase transition
in question, vanish in one of the phases and diverge in the other. 
Here we consider $R(L,g)=\langle W_-^2 \rangle$.
\
\item[(iii)] Perform a data collapse for flowgrams of $R(L,g)$,
by rescaling the linear system size, $L \to C(g)L$, where $C(g)$ is
a smooth and monotonically increasing function of the
coupling constant $g$. In the present case we have $C(g\to 0) \propto g$ \cite{Halperin}.
\end{enumerate}

A collapse of the rescaled flows within an interval $g\in [0,\, g_{\rm coll}]$
implies that the type of the transition within the interval remains the same,
and thus can be inferred by dealing with the $g=g_{\rm coll}$ point only. 
Since the  $g\to 0$ limit implies large spatial scales, and, therefore, model-independent
runaway renormalization flow pattern, the conclusions are universal.

\begin{figure}
\vspace*{-0.5cm}
\centerline{\includegraphics[scale=0.95, angle=0]{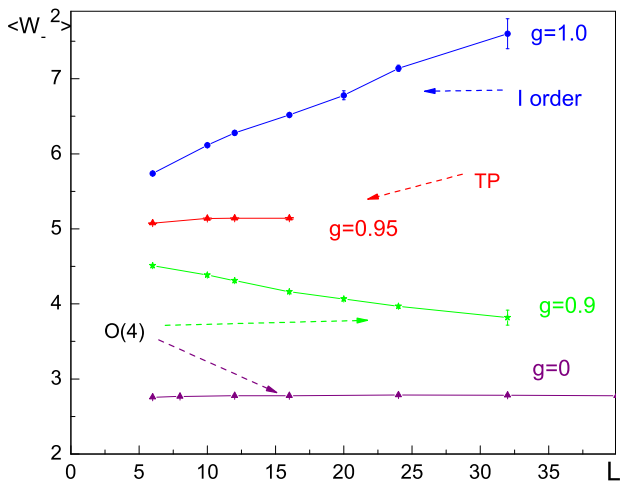}}
\vspace*{-1.0cm}
\caption{\small{(Color online) Flowgrams for the short-range
model. The lower horizontal line features the O(4) universality scaling behavior,
so that for $g<g_c \approx 0.95$ all flows are attracted to this
line. The upper horizontal line is the tricritical separatrix (marked as TP).
Above it, flows diverge due to the firs-order transition
detected by the bi-modal distribution of energy. }}\label{fig2}
\end{figure}

To have a reference comparison, we first simulated a short-range
analog of the NCCP$^1$ model (\ref{H_J}) with $V_{ij} =  g\delta_{ij}$.
The short-range model has a similar phase diagram, but with a second order phase transition for small $g$ and a first order one at large $g$. Figure \ref{fig2} clearly shows that the corresponding
flowgram cannot be collapsed on a single master curve by rescaling the length (shifting
the lines horizontally in logarithmical scale), and the separatrix at the tricritical point (TP) 
at $g\approx 0.95$ is clearly visible.

\begin{figure}[b]
\centerline{\includegraphics[scale=0.9, angle=0]{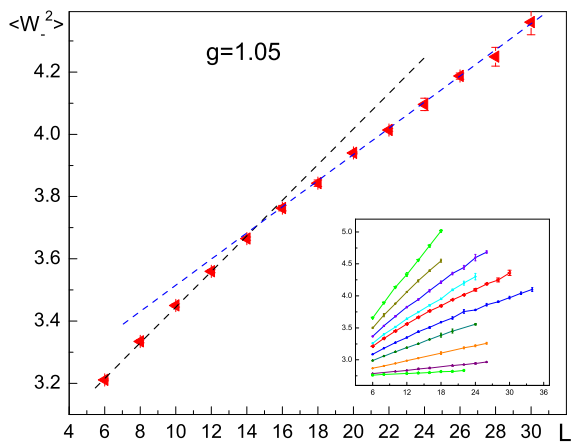}}
\vspace*{-1.0cm}
\caption{\small{(Color online) A typical flowgram of the gauge invariant superfluid stiffness in the NCCP$^1$ model.  The inset shows a fan of diverging flows for $0.125<g<1.4$}}\label{fig2a}
\end{figure}

Contrary to the short range model we find no such separatrix for the DCP action. 
As shown in Fig.~\ref{fig2a} the flows feature a fan of lines diverging with the 
system size and with the slope increasing with $g$ without any sign of a TP separatrix.

\begin{figure}
\vspace*{-0.5cm}
\centerline{\includegraphics[scale=0.95,angle=0]{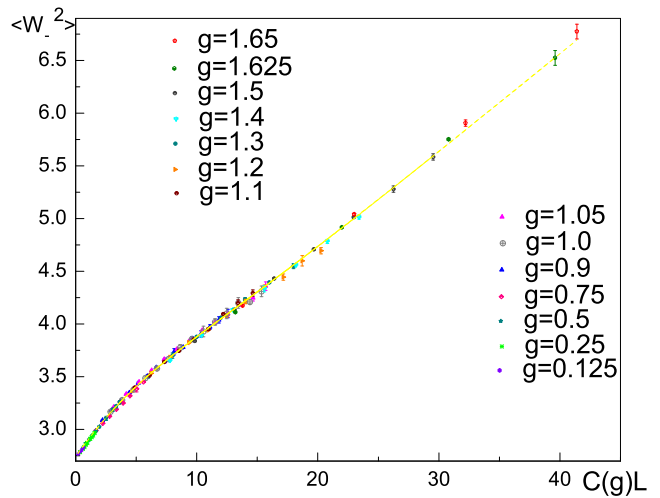}}
\vspace*{-0.6cm}
 \caption{\small{(Color online)
Data collapse for the NCCP$^1$ flows.  
The yellow line is a fit representing the master curve. The horizontal
axis is the scale reduced variable $C(g)L$ with $C(g)=(\exp(bg)-1)/(\exp(bg_1)-1),\, b=2.28 \pm 0.02$
and $g_1=1.3$. Error bars are shown for all data points.
}}\label{fig3}
\end{figure}
One can notice that the NCCP$^1$ flows exhibit a slope change, see Fig.~\ref{fig2a}
(also observed in Ref.~\cite{Wiese} for the {\it J-Q}-model) that might be interpreted
as a sign of the evolution towards a scale invariant behavior
$\langle W_-^2\rangle = {\rm const}$, possibly achieved at a large enough $L$.  The same
feature has been observed recently in Ref.~\cite{MV}, and caused the
authors to speculate that the NCCP$^1$ model features a line of
continuous transitions for $g<1.25$ \cite{note1}.
The crucial test, then, is to
see if the fan of the NCCP$^1$ lines can be collapsed on a single
master curve $\langle W_{-}^2\rangle = F(C(g)L)$, where
$C(g)$ describes the length-scale renormalization set by the coupling constant $g$.
As it turns out, the NCCP$^1$ flows collapse perfectly \cite{note2} in the whole region 
$0.125\leq g <1.65$ below the bicritical point $g_{bc}$ (see Fig.~\ref{fig3}).
The rescaling function $C(g)$ exhibits a linear behavior $C(g)\propto
g$ at small $g$ consistent with the runaway flow in the lowest-order
renormalization group analysis \cite{Halperin}.
This behavior all but rules out the existence of
the TP on the VBS-2SF line.

Though our conclusions directly contradict claims made in
Refs.~\cite{Sandvik2007,Melko2008,MV}, the primary data are in agreement. A data collapse
of the flowgram presented in the lower panel of Fig.~13 in Ref.~\cite{MV} shows
the same qualitative behavior as our Fig.~\ref{fig3}  \cite{comment}. 
We are also consistent with the conclusion reached in Ref.~\cite{Wiese} that the
slope change is an intermediate scale phenomenon and the 
N\`{e}el antiferromagnet to VBS transition in the {\it J-Q}-model violates
the scale invariance hypothesis as observed by the divergent flow of
$\langle W_{-}^2\rangle$.

\begin{figure}[t]
\vspace*{-0.5cm}
\centerline{\includegraphics[scale=0.9]{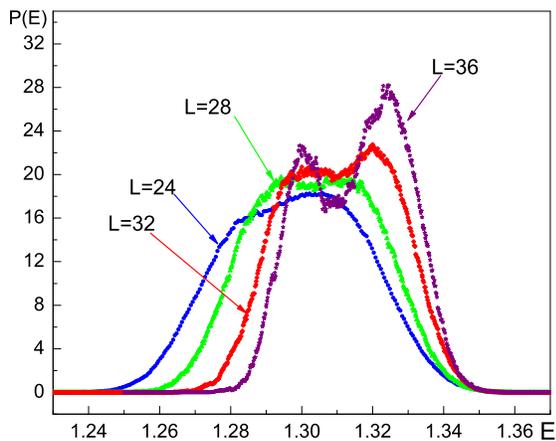}}
\vspace*{-1.0cm}
\caption{\small{(Color online) Evolution towards the bi-modal
energy distribution with increasing system size indicative of the
first-order deconfinement transition ($g=1.65$).
}}
\label{fig4}
\end{figure}

The flow collapse within an interval $g\in [0, g_{\rm coll}]$ does
not yet imply a first-order transition. What appears to be a
diverging behavior in Fig.~\ref{fig2a} might be just a reconstruction
of the flow from the O(4)-universality (at $g=0$) to a novel
DCP-universality at strong coupling. To complete the proof,
we have to determine the nature of the transition for $g=g_{\rm coll}$. 
In this parameter range the standard technique of detecting discontinuous transitions
by the bi-modal energy distribution becomes feasible. As shown in Fig.~\ref{fig4} 
a clear bi-modal distribution develops at $g=1.65$ which is below the bicritical point $g_{bc}$
and within the data collapse interval $[0, g_{\rm coll}]$.

This leaves us with the clear conclusion that the whole phase transition line for small
$g$ features a generic weak first-order transition identical to the one observed in the  U(1)$\times$U(1)
case. Driven by long-range interactions, this behavior develops on
length scales $\propto 1/g \to \infty$ for small $g$ and thus is universal.
It cannot be affected by microscopic variations of the NCCP$^{1}$ model
suggested in Ref.~\cite{MV} to suppress the paired (molecular) phase. 
 
We acknowledge useful discussions with O.~Motrunich,
A.~Vishwanath, L.~Balents and E.~Babaev. We thank the
Institut Henri Poincare-Centre Emile Borel and Nordita 
for hospitality and support in 2007. 
This work was supported by NSF under Grants Nos. PHY-0653135,
PHY-0653183 and CUNY grants 69191-0038, 80209-0914. We also
recognize the crucial role of the (super)computer clusters at
UMass, Typhon and Athena at CSI, and Hreidar at ETH.


\begin{thebibliography}{99}

\bibitem{Motrunich} O.I.~Motrunich and A.~Vishwanath,
Phys. Rev. {\bf B 70}, 075104 (2004).

\bibitem{dcp1} T.~Senthil, A.~Vishwanath, L.~Balents, S.~Sachdev,
and M.P.A.~Fisher, Science {\bf 303}, 1490 (2004).

\bibitem{dcp2} T.~Senthil, L.~Balents, S.~Sachdev, A.~Vishwanath,
and M.P.A.~Fisher, Phys. Rev. {\bf B 70} 144407 (2004).

\bibitem{Babaev}  E.~Babaev, Nucl. Phys. {\bf B 686}, 397 (2004);
E.~ Babaev, A.~ Sudbo, N. W.~ Ashcroft, Nature {\bf 431}, 666 (2004));
J.~Smiseth, E.~Sm{\o}rgrav, E.~Babaev, and A.~Sudb{\o},
Phys. Rev. B {\bf 71}, 214509 (2005).

\bibitem{Halperin} B.I.~Halperin, T.C.~Lubensky, and S.-K.~Ma,
Phys. Rev. Lett. {\bf 32}, 292 (1974); E.~Br$\acute{\rm e}$zin,
J.C.~Le~Guillou, and J.~Zinn-Justin, Phys. Rev. {\bf B 10}, 892
(1974); J.-H.~Chen, T.C.~Lubensky, and D.~Nelson, Phys. Rev. {\bf
B 17}, 4274 (1978).

\bibitem{Sachdev} L. Balents, L. Bartosch, A. Burkov, S. Sachdev, and K. Sengupta,
Phys. Rev. B 71 144509 (2005); ibid 144508 (2005).

\bibitem{duality}  M.~Peskin, Ann. Phys. (N.Y.) {\bf 113}, 122  (1978);
P.R.~Thomas and M.~Stone, Nucl. Phys. {\bf B 144}, 513 (1978);
C.~Dasgupta and B.I.~Halperin, Phys. Rev. Lett. {\bf 47}, 1556
(1981).


\bibitem{Sandvik2002} A.W.~Sandvik, S.~Daul, R.R.P.~Singh, and D.J.~Scalapino, Phys.
Rev. Lett. {\bf 89}, 247201 (2002).

\bibitem{weak_first} A.~Kuklov, N.~Prokof'ev, and B.~Svistunov,  Phys.
Rev. Lett. {\bf 93}, 230402 (2004).

\bibitem{prog_theor} A.~Kuklov, N.V.~Prokof'ev, and B.V.~Svistunov,
Prog. of Theor. Phys. Suppl. {\bf 160}, 337 (2005).



\bibitem{flowgram}  A.~Kuklov, N.~Prokof'ev, B.~Svistunov, and M.~Troyer, Ann. Phys. (N.Y.) {\bf 321}, 1602 (2006).

\bibitem{Sandvik2006} A.W.~Sandvik and R.G.~Melko, Ann. Phys. (N.Y.) {\bf 321}, 1651
(2006).

\bibitem{Sudbo2006} S.~Kragset, E.~Sm{\o}rgrav, J.~Hove, F.S.~Nogueira, and A.~Sudb{\o},
Phys. Rev. Lett. {\bf 97}, 247201 (2006).

\bibitem{Sandvik2007} A.W.~Sandvik, Phys. Rev. Lett. {\bf 98}, 227202 (2007).

\bibitem{Melko2008} R.G.~Melko and R.K.~Kaul,
Phys. Rev. Lett. {\bf 100}, 017203 (2008).

\bibitem{MV} O.I.~Motrunich and A.~Vishwanath, arXiv:0805.1494.

\bibitem{Wiese} F.-J.~Jiang, M.~Nyfeler, S.~Chandrasekharan, and U.-J.~Wiese,
arXiv:0710.3926.

\bibitem{su2} Preliminary results have been
announced: A.~Kuklov, M.~Matsumoto, N.~Prokof'ev, B.~Svistunov, and
M.~Troyer, Bull. Am. Phys. Soc. {\bf 53}, S12.00006 (2008); and a presentation  by A.~Kuklov at the {\it Quantum Fluids} workshop
(Nordita, Stockholm, August 15 - September 30, 2007)
http://www.nordita.org/$\sim$qf2007/kuklov.pdf .

\bibitem{note} See Ref.~\cite{Babaev} for discussions of 2d as well as 3d field induced paired phases in two-component superconductors. 


\bibitem{note1} The interaction constant $K$
in Ref.~\cite{MV} is defined as $K=1/(4g)$.

\bibitem{note2} A flow collapse
is meaningful even when the collapsing lines $R(L)$ are
relatively short
and reminiscent of straight lines: a straight line is described
by two independent parameters, while the rescaling procedure has only one
degree of freedom of shifting the line horizontally in logarithmical scale. The master curve may significantly deviate from a straight line and
prove indispensable for understanding the global character of the flow
and difficulties with the finite-size scaling in specific models.

\bibitem{comment} A.~Kuklov, M.~Matsumoto, N.~Prokof'ev, B.~Svistunov, and
M.~Troyer, ArXive:0805.2578.
 

\end{thebibliography}
\end{document}